# Magnetic irreversibility and pinning force density in the Mo$_{100-x}$Re$_x$ alloy superconductors

**Shyam Sundar[1, 2], M. K. Chattopadhyay[1, 2]\*, L. S. Sharath Chandra[2] and S. B. Roy[1, 2]**

[1]*Homi Bhabha National Institute, Raja Ramanna Center for Advanced Technology, Indore 452 013, India*

[2]*Magnetic & Superconducting Materials Section, Raja Ramanna Center for Advanced Technology, Indore 452 013, India*

## Abstract

We have measured the isothermal field dependence of magnetization of the Mo$_{100-x}$Re$_x$ ($15 \leq x \leq 48$) alloys, and have estimated the critical current and pinning force density from these measurements. We have performed structural characterization of the above alloys using standard techniques, and analyzed the field dependence of critical current and pinning force density using existing theories. Our results indicate that dislocation networks and point defects like voids and interstitial imperfections are the main flux line pinning centres in the Mo$_{100-x}$Re$_x$ alloys in the intermediate fields, i.e., in the "small bundle" flux line pinning regime. In this regime, the critical current density is also quite robust against increasing magnetic field. In still higher fields, the critical current density is affected by flux creep. In the low field regime, on the other hand, the pinning of the flux lines seems to be influenced by the presence of two superconducting energy gaps in the Mo$_{100-x}$Re$_x$ alloys. This modifies the field dependence of critical current density, and also seems to contribute to the asymmetry in the magnetic irreversibility exhibited by the isothermal field dependence of magnetization.

# 1. Introduction

The $Mo_{100-x}Re_x$ alloy superconductors are an interesting system where the normal to superconducting transition temperature ($T_c$) of the alloys are higher than both Mo and Re [1]. The $T_c$ of many of these alloys are relatively high (near 12 K) [1] which is important from the application point of view. The $Mo_{100-x}Re_x$ alloys are also known to have good mechanical properties and they are widely used in the aerospace and defense industries [2- 4]. In view of the scope of technological applications, superconducting solenoids made out of the $Mo_{100-x}Re_x$ alloys have been tested in past [5], and the field ($H$) hysteresis of magnetization ($M$) of the $Mo_{66}Re_{34}$ alloy has also been studied [6- 8]. This latter study has revealed that the dislocation networks are the main source of flux line pinning in the $Mo_{66}Re_{34}$ alloy. However, a comprehensive study on the $H$ dependence of critical current density ($J_c$) as a function of Re-content is not found in literature. It is also observed that in spite of substantial information available in literature [1- 8], the superconducting properties of the $Mo_{100-x}Re_x$ alloys are still not very well-studied and only our very recent study indicated the presence the two superconducting energy gaps in this alloy system [9]. This observation of two superconducting energy gaps now raises a natural question: how does this phenomenon influence the various superconducting properties of the system, especially in the presence of magnetic fields?

In the present work, we have synthesized as-cast and annealed $Mo_{100-x}Re_x$ alloys of five different compositions, covering a wide range of Re-concentration, and have studied these alloys with the help of magnetization experiments. We have estimated the $J_c$ and the pinning force density ($F_p$) of these alloys from the $M(H)$ curves, and have analyzed the nature of the magnetic irreversibility and the field dependence of $J_c$ and $F_p$ with the help of existing theories. Our analysis reveals the flux line pinning mechanisms available in this alloy system in different fields. Our results also indicate that the $J_c$ of these alloys in the low field regime is indeed influenced by the presence of two superconducting energy gaps [9], and this does give rise to substantial changes in the $F_p(H)$ in low $H$ which seems to influence the symmetry of the $M(H)$ curves between increasing and decreasing $H$.

## 2. Experimental

Polycrystalline samples of $Mo_{100-x}Re_x$ (x = 15, 20, 25, 40, and 48) alloys were synthesized by melting high purity (99.95+ %) Mo and Re (ACI Alloys, USA) taken in atomic proportions. The melting was performed under high purity Ar atmosphere in an arc-melting furnace. The sample was flipped and re-melted six times to ensure the homogeneity. The loss of mass during the melting procedure was less than 0.1 %. The as-cast buttons were cut in two halves with the help of a spark wire cutter, and one of the halves was wrapped in Ta foil and sealed in quartz ampoule in Ar atmosphere for annealing. The annealing was done at 1250 °C for 20 hours in Ar atmosphere, which was followed by slow cooling down to the room temperature [10]. The samples were cut in different shapes and sizes with the help of diamond wheel cutter and spark wire cutter for different experimental studies.

The structural characterization of the present $Mo_{100-x}Re_x$ alloy samples was done with the help of X-Ray Diffraction (XRD), optical metallography, and energy dispersive x-ray (EDX) measurements. The XRD experiments were performed in a standard diffractometer (Rigaku Corporation, Japan: Geigerflex model) using Cu $K\alpha$ ($\lambda$ = 1.5418 Å) radiation. For optical metallography, small pieces of samples were mounted on moulds prepared from epoxy (Resin + Hardener) and then ground, to make them flat, with the help of silicon carbide abrasive papers. Properly ground moulds were then polished to 0.5 $\mu$m roughness with the help of diamond paste. Polishing of the sample was done by holding the moulds on rotating Nylon cloth. An etchant solution of $H_2O_2$ (50%) and $NH_4OH$ (50%) were used for the visualization of the microstructure of the polished samples. The metallographic characterization of the samples was done using a high power optical microscope (Leica DMI 5000M). The chemical compositions of the samples were experimentally confirmed through EDX analysis performed using a Philips XL-30pc machine. Within the experimental resolution of the EDX measurements, the chemical compositions were not found to vary over the bulk of the samples.

The measurement of $M$ as functions of temperature ($T$) and $H$ were performed using a liquid He based Vibrating Sample Magnetometer (VSM; Quantum Design, USA) and a closed cycle refrigerator based MPMS-3 SQUID VSM (Quantum Design, USA). In the VSM, the isothermal $M$ versus $H$ measurements were performed while varying $\mu_0 H$ sequentially from 0 to 8 T, from 8T to -8 T, and then from -8 T to 8 T, starting from an initial zero-field-cooled (ZFC) state. The $M(H)$ curves obtained in the last two runs of the above measurement, i.e., while varying $\mu_0 H$ sequentially from 8T to -8 T, and then from -8 T to 8 T (the upper and lower envelope magnetization curves respectively) will henceforth be referred to as $M^-$ and $M^+$ respectively. The protocol for the isothermal $M(H)$ measurements performed in the MPMS-3 SQUID VSM was kept same as that of the VSM. Only, the maximum field available in MPMS-3 SQUID VSM is ±7 T. After measuring $M^+$ and $M^-$ isothermally, the MPMS-3 SQUID VSM was used for drawing minor hysteresis loops (MHLs). Two kinds of MHLs were constructed for the present study. In the first kind, the MHL was initiated by starting $M(H)$ measurements while decreasing $H$ from a field value on the $M^+$ curve. The MHL was continued till it merged with the $M^-$ curve. For the second kind of MHL, the sample was cooled in constant $H$ from 50 K down to 2 K (FC protocol), and the then $M(H)$ was measured at 2 K while increasing (or decreasing) the $H$ till the MHL merged with the $M^+$ curve (or the $M^-$ curve). These MHLs will be referred to as the FC MHLs. For every value of $H$ in which the FC state is prepared, two FC MHLs were drawn with increasing and decreasing $H$, and they gave two values of $M$ where the FC MHLs reached the $M^+$ and $M^-$ curves. Apart from the above measurements, the magnetic relaxation experiments were also performed in the MPMS-3 SQUID VSM. For these experiments $M$ was measured for about 90 minutes at constant $T$ (2 K) and $H$, starting from different $H$-values on the $M^+$ and $M^-$ curves. The choice of the VSM or the MPMS-3 SQUID VSM for the above measurements is based on the availability of liquid He and the availability of the equipment at the time of the measurement.

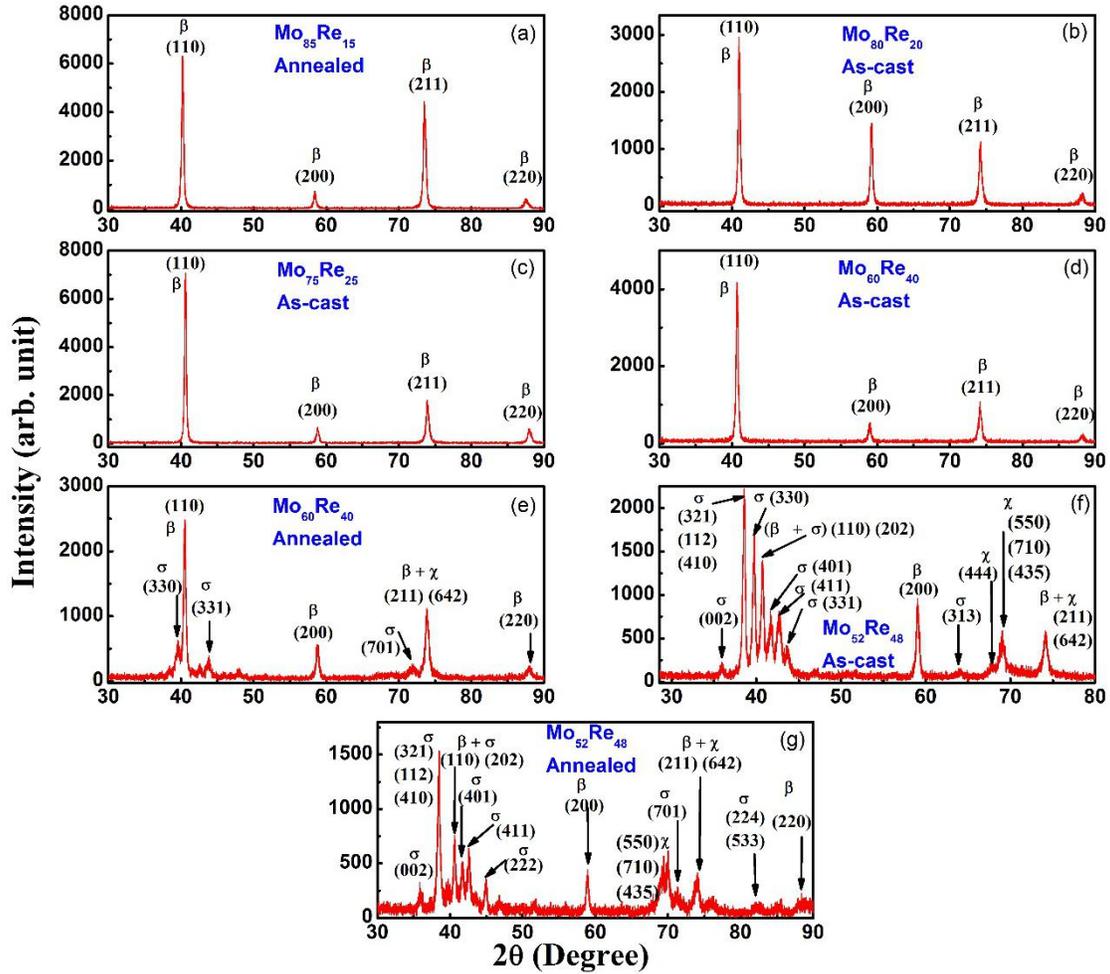

**FIG. 1.** XRD patterns of the $Mo_{100-x}Re_x$ alloys indicating the presence of the $\beta$, $\sigma$ and $\chi$ phases. Only the indices corresponding to the prominent peaks are shown here.

## 3. Results and discussion

### 3.1. Structural characterization, grain size, and the superconducting transition temperature

Fig. 1 shows the XRD patterns for the annealed $Mo_{85}Re_{15}$ alloy, as-cast $Mo_{80}Re_{20}$ and $Mo_{75}Re_{25}$ alloys, and the annealed and as-cast $Mo_{60}Re_{40}$ and $Mo_{52}Re_{48}$ alloys. The XRD patterns indicate that the annealed $Mo_{85}Re_{15}$ alloy, and the as-cast $Mo_{80}Re_{20}$, $Mo_{75}Re_{25}$ and $Mo_{60}Re_{40}$ alloys are single phase, and have formed in the body centred cubic (bcc) $\beta$ phase structure. On the other hand, the XRD patterns of the annealed $Mo_{60}Re_{40}$ alloy, and the annealed and as-cast $Mo_{52}Re_{48}$ alloys, indicate the presence of the $\beta$, $\sigma$ (tetragonal), and $\chi$ (complex bcc) phases in these samples (see table I).

TABLE 1. Metallurgical phases present in the $Mo_{100-x}Re_x$ alloys along with the approximate phase fractions (estimated from the XRD patterns), and the $T_c$.

| Alloy Compositions | Metallurgical Phases Present | $T_c$ (K) | $\mu_0 H_{C1}$ (T) at 2K |
|---|---|---|---|
| $Mo_{85}Re_{15}$, Annealed | $\beta$ | 6.89 | 0.0430 |
| $Mo_{80}Re_{20}$ As-cast | $\beta$ | 8.63 | 0.0557 |
| $Mo_{75}Re_{25}$ As-cast | $\beta$ | 9.85 | 0.0675 |
| $Mo_{60}Re_{40}$ As-cast | $\beta$ | 12.78 | 0.0815 |
| $Mo_{60}Re_{40}$ Annealed | $\beta$ (75 %), $\sigma$ (20 %), $\chi$ (5 %) | 12.44 | 0.0900 |
| $Mo_{52}Re_{48}$ As-cast | $\beta$ (50 %), $\sigma$ (40 %), $\chi$ (10 %) | 12.68 | 0.0548 |
| $Mo_{52}Re_{48}$ Annealed | $\beta$ (50 %), $\sigma$ (40 %), $\chi$ (10 %) | 11.85 | 0.0438 |

Fig. 2(a) shows a selected optical micrograph of the annealed $Mo_{85}Re_{15}$ alloy. Large grains with varying sizes and a strong network of dislocations [6- 8] are observed in this micrograph. We have performed optical metallography experiments on different portions of each of the present samples. Only a few representative micrographs are shown in Fig. 2. In general, the grains of the present $Mo_{100-x}Re_x$ alloys are found to be very large in size. In the single phase $Mo_{100-x}Re_x$ alloys used in the present work (see table I), e.g., the annealed $Mo_{85}Re_{15}$ alloy, the as-cast $Mo_{80}Re_{20}$, $Mo_{75}Re_{25}$, and $Mo_{60}Re_{40}$ alloys, the grains are found to have large variation of size over the sample surface. Overall, the grain sizes in these alloys typically vary from 100 to 600 $\mu$m, with some grains as large as 2000 $\mu$m. Fig. 2(b) shows an optical micrograph of the as-cast $Mo_{80}Re_{20}$ alloy in a magnified scale. This micrograph, obtained after very careful polishing and etching, provides a clear view of the dislocation network in the form of lining up of the etch pits. Such dislocation network is visible in all the single phase $Mo_{100-x}Re_x$ alloys described above.

The observation of dislocation networks in the present $Mo_{100-x}Re_x$ alloys is consistent with the existing literature [6- 8]. Fig. 2(c) shows an optical micrograph of the annealed $Mo_{60}Re_{40}$ alloy, which contains $\beta$, $\sigma$ and $\chi$ phases. Apart from the grain boundaries visible in this micrograph, it appears that the $\sigma$ and $\chi$ phases precipitate around the dislocation networks in this alloy. The precipitation of the $\sigma$ phase along the dislocation networks in the $Mo_{100-x}Re_x$ alloys have been reported previously [8]. The average grain size in the annealed $Mo_{60}Re_{40}$ alloy is 50-150 $\mu$m. The average grain size in the (multi-phase) $Mo_{52}Re_{48}$ as-cast alloy [Fig. 2(d)] is found to be 20- 100 $\mu$m. No indication of dislocation network is found in the optical micrographs of the as-cast $Mo_{52}Re_{48}$ alloy or its annealed counterpart (not shown here). In general, the average grain sizes in the multi-phase $Mo_{100-x}Re_x$ alloys are found to be much smaller and more evenly distributed as compared to the single phase alloys.

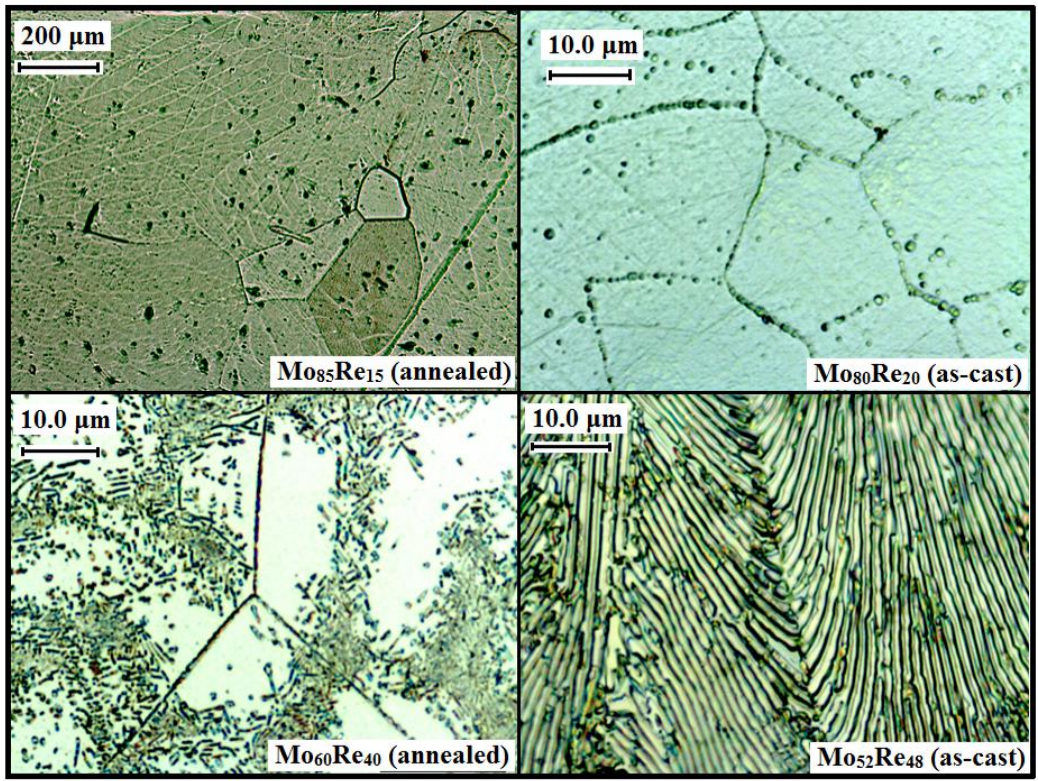

**FIG. 2.** Optical micrographs for selected $Mo_{100-x}Re_x$ alloys. While the panels (a) and (b) are for the single ($\beta$) phase samples, the panels (c) and (d) are for multi-phase samples containing $\beta$, $\sigma$ and $\chi$ phases.

The $T_c$ values for the present samples are shown in table 1. These values were obtained by finding the $T$-value at which the $M(T)$ curves (measured in 10 mT magnetic field, not presented here) start to drop downwards towards a negative value with decreasing $T$, and these values are consistent with the literature reports [11]. We have found that the $T_c$ values for the annealed and as-cast $Mo_{85}Re_{15}$ alloys are not different, and we have used the annealed $Mo_{85}Re_{15}$ alloy only for the present study. For the other alloys, the $T_c$ values for the as-cast alloys are slightly higher than their annealed counterparts.

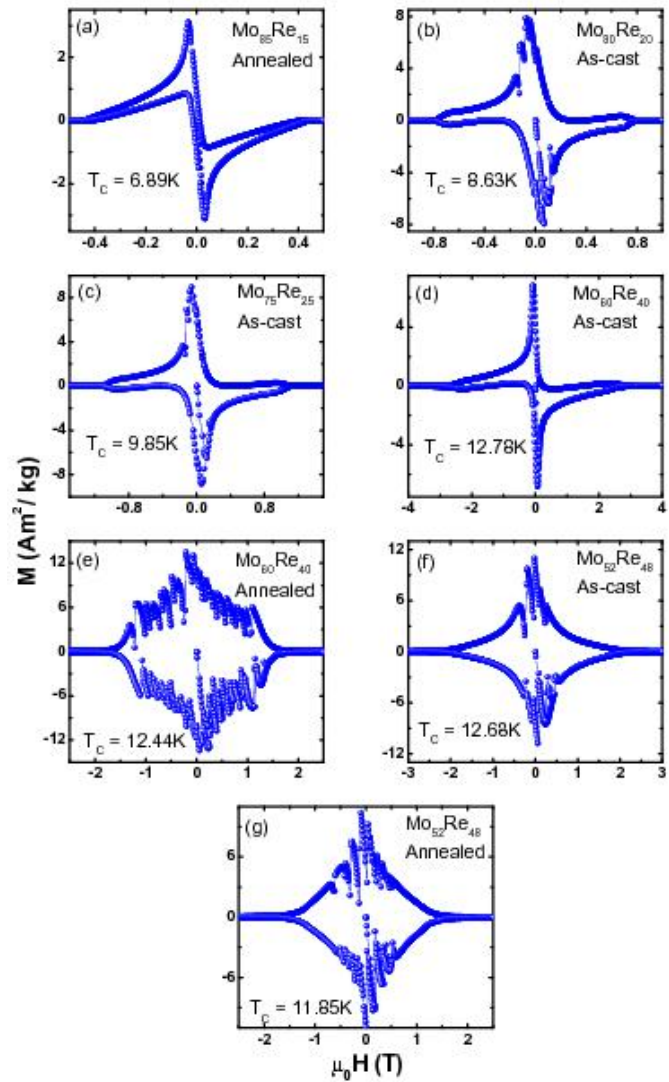

**FIG. 3.** Field dependence of magnetization of the $Mo_{100-x}Re_x$ alloys at 2K, measured in the VSM.

## 3.2. The field dependence of *M*, critical current density and pinning force density

Fig. 3 shows the isothermal *M(H)* curves for the present alloys at 2 K depicting the irreversible magnetic behaviour of the present alloys. The major part of this magnetic irreversibility reduces quite rapidly with increasing *H* for all the present alloys. A small amount of irreversibility between increasing and decreasing *H*, however, is sustained in all the present alloys up to high *H*. For the annealed $Mo_{85}Re_{15}$ alloy, this magnetic irreversibility disappears at $\mu_0 H = 1$ T at 2 K. The field at which this small magnetic irreversibility disappears completely, is enhanced considerably with increasing Re content. For the as-cast $Mo_{60}Re_{40}$ alloy, at 2 K, a tailing magnetic hysteresis remains up to 5.6 T. For the annealed $Mo_{60}Re_{40}$ alloy, this tailing magnetic hysteresis is observed even in 8 T (at 2K). For the annealed and as-cast $Mo_{52}Re_{48}$ alloys the phenomenon is sustained up to 7 T (at 2 K). This tailing magnetic hysteresis, which is not visible in Fig. 3, and is observable only in a highly magnified scale, may be related to the presence of the third critical field $H_{c3}$ reported in the $Mo_{100-x}Re_x$ alloys [10]. Fig. 3 also shows that there is considerable asymmetry between the $M^+$ and $M^-$ curves for the single phase $Mo_{100-x}Re_x$ alloys (see table I). The major portions of the $M^-$ curves for these alloys are negative for positive *H*-values, which leads to asymmetry in *M(H)*, and the very narrow width of the hysteresis in most fields depicted in Figs. 3(a) to (d). The magnetic hysteresis for these alloys is significant only in very low *H*. Such asymmetric hysteresis can sometimes appear because of surface barrier effects [12- 14]. It is already known that the presence of surface barrier effect in the superconductors may be confirmed with the help of the MHL technique [12]. We have, however, found that for the present alloys, the MHL initiated from the $M^+$ curve touches the $M^-$ curve after showing a rounding-off behaviour [12]. This is shown in Fig. 4 for the $Mo_{75}Re_{25}$ as-cast alloy. Similar curves exist for the other alloys as well, and they indicate that the surface barrier effect may not be significant in the present samples. Narrow isothermal *M(H)*-hysteresis curves similar to the present alloys have been observed both in the polycrystalline [15] and single crystal $MgB_2$ samples [16]. In these materials, the narrow width of the hysteresis, is thought to be arising due to weak pinning [15] and the lack of defects (flux line pinning centres) [16]. In such a case, the $M^-$ curve lies close to the equilibrium magnetization ($M_{Eq}$) and thus exhibits negative values. Fig. 3 also shows that the $M(H)$ curves for the annealed $Mo_{60}Re_{40}$ alloy and the as-cast and annealed $Mo_{52}Re_{48}$ alloys do not exhibit such a pronounced asymmetry. On the other hand, the $M(H)$ curves for these multi-phase samples show the signatures of flux jump effects.

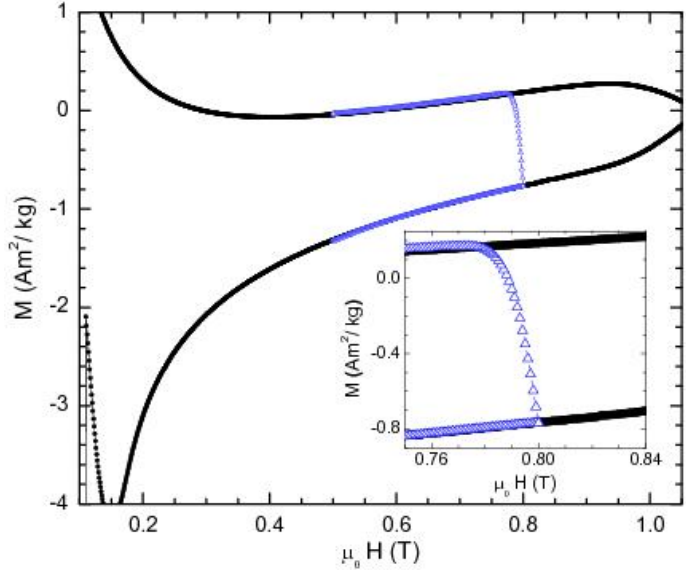

**FIG. 4.** Field dependence of magnetization of the as-cast $Mo_{75}Re_{25}$ alloy at 2 K. The black dots represent the envelope magnetization curves, and the (blue) open triangles represent the MHL initiated from the lower envelope magnetization curve. The inset shows a magnified version of the relevant portion of the curves. These experiments were performed in MPMS-3 SQUID VSM.

Fig. 5 shows the $J_c(H)$ curves for the present alloys, estimated from the irreversibility in $M(H)$. Only the curves for a few selected temperatures are shown here. The $J_c$ values were estimated with the help of Bean model [17], using the formula $J_c = \Delta M\ [a_2(1-a_2/3a_1)]^{-1}$, where $\Delta M$ is the difference in $M$ (for a particular $H$) between the $M^+$ and $M^-$ curves [18, 19]. The parameters $2a_1$ and $2a_2$ ($a_1 > a_2$) are the dimensions of the rectangular samples used in the present measurements, in directions normal to the applied $H$. The $J_c(H)$ curves shown in fig. 5 are initiated from a field above the lower critical field $H_{c1}$ so as to avoid the field regime where the flux-penetration in the sample is not uniform. The method of determination of $H_{c1}$ for the present alloys has been discussed in our previous work [9]. The $\mu_0 H_{c1}$ values for the present $Mo_{100-x}Re_x$ alloys are shown in table 1. For the samples exhibiting flux jumps, we have plotted the $J_c(H)$ curves in fields still higher than $H_{c1}$, in the regime where these jumps are not observed. It may be observed in fig. 3 that the multi-phase $Mo_{100-x}Re_x$ alloys exhibit higher values of $\Delta M$ over a larger $H$-regime as compared to the single ($\beta$) phase alloys. This indicates that the $J_c$ and $F_p$ values for the multi-phase $Mo_{100-x}Re_x$ alloys are higher than the single ($\beta$) phase alloys. Incidentally, these multi-phase alloys also exhibit higher $T_c$ values (see table 1) and smaller grain size. Among the multi-phase alloys, the largest $\Delta M$ is observed in the annealed $Mo_{60}Re_{40}$ alloy [see fig. 3(e)]. We have however not estimated the $J_c$ for the annealed $Mo_{60}Re_{40}$ alloy because of the excessive flux jump effects in this alloy. It may however be noted that stronger flux jump effects also indicate higher $F_p$. We have also extracted the $T$-dependence of $J_c$ of the $Mo_{100-x}Re_x$ alloys, for different $H$-values, from the isothermal $J_c(H)$ curves. All these $J_c(T)$ curves (not shown here) fall nearly linearly with increasing $T$, with a slight tendency of bending further downwards at higher $T$ indicating the absence of weak links in the present alloys. In the presence of weak links, these curves are expected to exhibit a tendency of bending upwards [20]. The formation of weak links (e. g. between the grains) is somewhat like the formation of S-N-S type junctions, and this leads to a upward curvature in the $J_c(T)$ curves (see Ref. [21] and references therein). It is therefore reasonable to assume that the current flow in the present samples is not fragmented because of the presence of grain boundaries or any other weak link.

It is observed in Fig. 5 that the $J_c(H)$ curves for the single ($\beta$) phase $Mo_{100-x}Re_x$ alloys (see table I) used in the present study may be qualitatively divided into three regimes: A low $H$ regime where the $J_c$ drops sharply with applied $H$, an intermediate $H$ regime where the $J_c$ varies slowly with increasing $H$, and a higher $H$ regime where the $J_c$ again drops rapidly with increasing $H$. While the $J_c(H)$ curves for the multi-phase $Mo_{100-x}Re_x$ alloys in the low $H$ regime could not be studied because of the flux jumps, the $J_c(H)$ behaviour for these alloys in the high $H$ regime seems to be quite similar to the single phase alloys. Both Figs. 3 and 5 provide hints of the existence of a peak effect (PE) in the higher field regimes of the $M(H)$ and $J_c(H)$ curves of the single phase $Mo_{100-x}Re_x$ alloys. However, it has been reported earlier that $M$ in the PE regime of the $Mo_{100-x}Re_x$ alloys does not exhibit any anomalous behaviour and can be explained using conventional theories (see Ref. [22] and references therein).

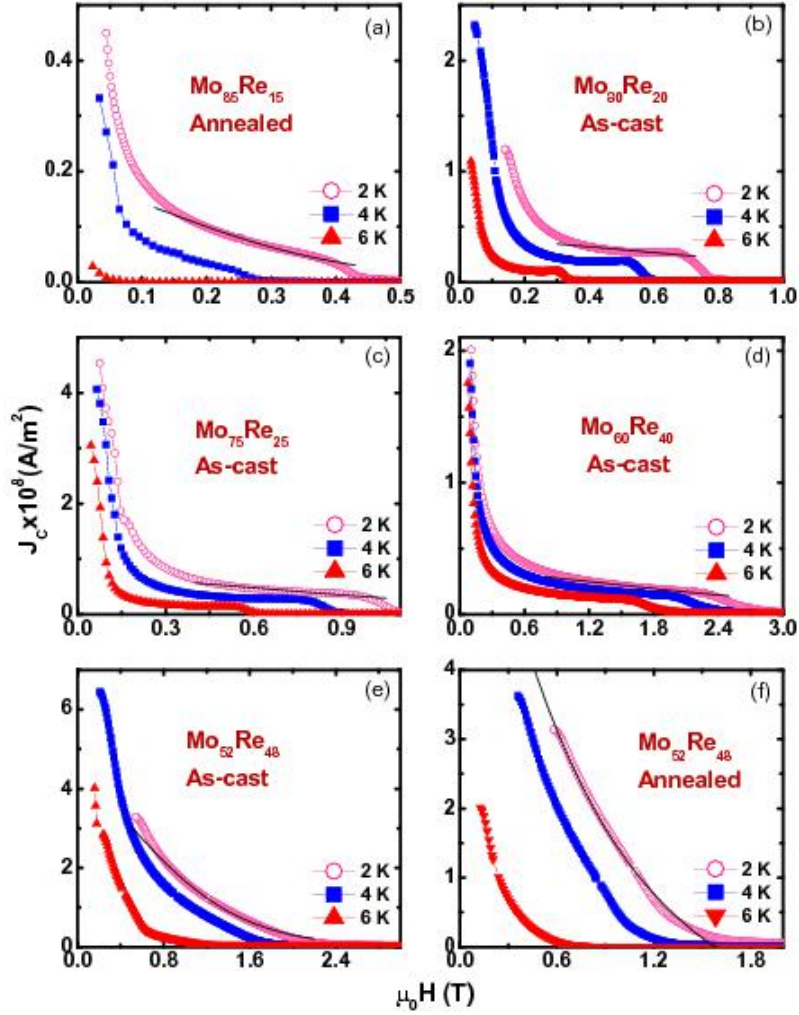

**FIG. 5.** Field dependence of critical current density of the $Mo_{100-x}Re_x$ alloys estimated from the isothermal $M(H)$ curves at selected temperatures.

We now look into the $J_c(H)$ behaviour of the $Mo_{100-x}Re_x$ alloys in the perspective of the existing theories [23]. According to the collective pinning models for the type-II superconductors, the separation ($a_0$) between the flux lines is very large in low $H$ so that the interaction between the flux lines is negligible. In this regime individual flux lines are pinned by the available pinning centres and the elasticity of the flux lines is not relevant. As the $H$ is increased, the interaction between the flux lines become significant, and then the elasticity (or the lack of rigidity) of the flux lines becomes more and more important for the nature of flux line pinning. The flux lines are displaced under the influence of the Lorentz force due the applied *H*. As long as the displacement of the flux lines is less than the coherence length ($\xi$), the flux lines can be considered rigid. For slightly higher *H*, the flux lines tend to shear, and the individual flux lines are then collectively pinned by several pinning centers. This is the so called "single vortex collective pinning regime". Up to this regime, the *J$_c$(H)* is expected to remain constant [23]. The regime of constant *J$_c$(H)* is expected to be quite narrow according to the collective pinning models. With increasing *H*, the flux line lattice undergo further displacement and it adjusts with the disorder (or pinning) potentials through shear and tilt deformations [23]. When the displacement of the shearing flux lines is of the order of *a$_0$*, the flux lines start interacting strongly with each other. Moreover, when the displacement is more than *a$_0$* the average pinning potential experienced by the flux line lattice remains approximately the same even after the displacement. The effect of individual disorder potentials becomes weak for displacements of the order of *a$_0$*. Bundles of flux lines are then collectively pinned by a spatial average of the pinning potentials. The size and aspect ratio of these flux line bundles are functions of *H* (and hence *a$_0$*) and the relevant elastic constants of the flux line lattice. It has been shown that there are elastic length scales that determine the volume (elongated in the direction of *H*) of these flux line bundles, and it grows rapidly with increasing *H* [23]. If the transverse elastic length-scale is less than the penetration depth, then the flux line bundles are considered "small". On the other hand, the flux line bundles are considered "large" if the penetration depth is smaller. In the so called "small bundle" flux line pinning regime, the *J$_c$(H)* is expected to be proportional to $\exp\left[-(\mu_0 H)^{\frac{3}{2}}\right]$ [23]. In fig. 5, this regime is indicated by the fitted line for different Mo$_{100-x}$Re$_x$ alloys at 2 K. The sharp decrease in *J$_c$(H)* in the Mo$_{100-x}$Re$_x$ alloys with increasing *H* in the lowest *H* regime, which extends over quite a broad *H*-range, is contrary to the normal expectation in the individual (or "single vortex") flux line pinning regime [23].

The pinning force density in the present Mo$_{100-x}$Re$_x$ alloys was estimated using the relation *F$_p$* = *J$_c$μ$_0$H* [24], and are presented in Fig. 6. For the single phase alloys, the *F$_p$(H)* curves are shown for *H* values greater than or equal to 1.5 times of $H_{c1}$ (the rationale for this choice has been explained below, in connection with Fig. 7) to ensure that the data points are free from errors due to non-uniform flux penetration. For the multi-phase alloys, the $F_p(H)$ curves are plotted for the $H$-regime with no flux jump. For the single phase $Mo_{100-x}Re_x$ alloys [Figs. 6(a)- (d)], the $F_p(H)$ curves exhibit a tendency of rising upwards in low $H$. Since $F_p = 0$ at $H = 0$, the above tendency of rising upwards in low $H$ actually indicates the presence of a low $H$ maximum in the $F_p(H)$ curves, and thus a double peak behaviour of these curves. The $F_p(H)$ curves for the multi-phase $Mo_{100-x}Re_x$ alloys, on the other hand, appear to exhibit a single peak behaviour [figs. 6(e) and (f)].

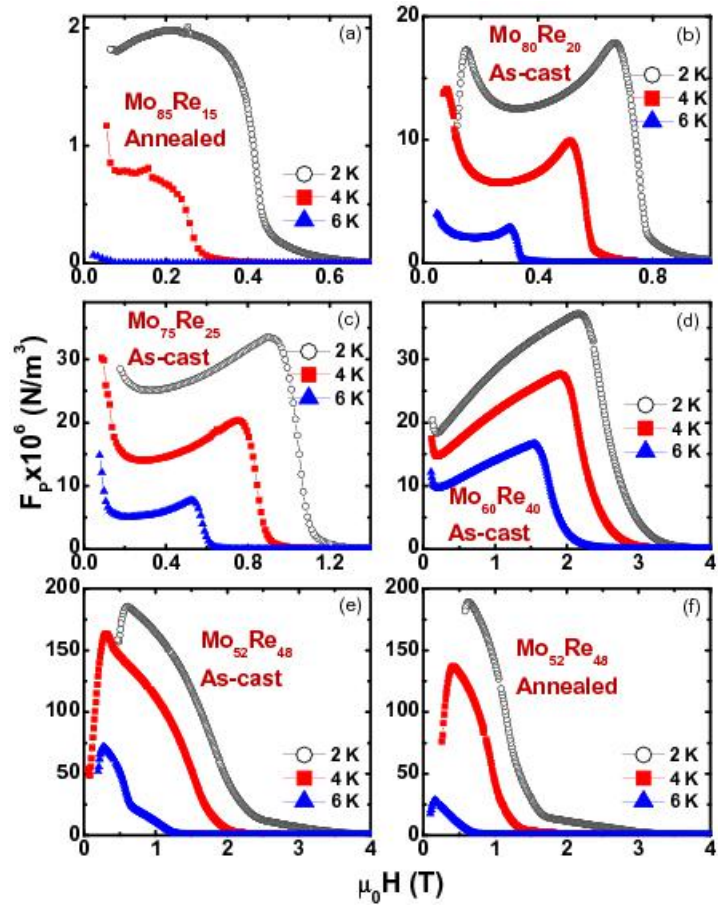

**FIG. 6.** Field dependence of pinning force density in the $Mo_{100-x}Re_x$ alloys.

Here we have estimated $J_c$ and $F_p$ from the isothermal $M(H)$ curves using the most common method described above. However, it has been reported in literature that this method becomes erroneous when the effect of the equilibrium (reversible) magnetization $M_{Eq}$ is significant as compared to the magnetic irreversibility. In such a case the $J_c$ and $F_p$ need to be estimated as a function of $B$, where, $B = \mu_0 (H + M_{Eq})$ [25]. Experimentally, $M_{Eq}$ is commonly determined as $M_{Eq} = (M^+ + M^-)/2$ [26]. However, when the magnitude of $M_{Eq}$ is significant as compared to $\Delta M$, a more accurate method of determination of $M_{Eq}$ is to use the FC MHL technique [26] described earlier. In this technique, the average of the two $M$-values where the FC MHLs touch the $M^+$ and $M^-$ curves gives the correct $M_{Eq}$ [26]. In Fig. 7(a), the $M_{Eq}$ obtained at different $H$-values using the FC MHL technique [(red) stars] is compared with the $M_{Eq} = (M^+ + M^-)/2$ curve [empty (blue) squares]. The $M_{Eq}$ values obtained in both the methods described above match very well in high $H$, and they differ appreciably only for $H < (1.5\ H_{c1})$. Similar results exist for our other $Mo_{100-x}Re_x$ alloys as well. In Fig. 7(b) we show a comparison (at 2 K) between the $F_p$ versus $\mu_0 H$ and $F_p$ versus $B$ curves for the annealed $Mo_{85}Re_{15}$ alloy where $B = \mu_0 (H + M_{Eq})$. The data below $H = (1.5\ H_{c1})$ are not shown in this figure because of the uncertainties discussed above. For the rest of the $H$ regimes, the two curves are almost identical. Therefore, to keep the method simple, we will continue to work with the $F_p$ versus $\mu_0 H$ curves of the present alloys and analyze the results for $H > 1.5 H_{c1}$ (or still higher in the cases where the flux jumps are present).

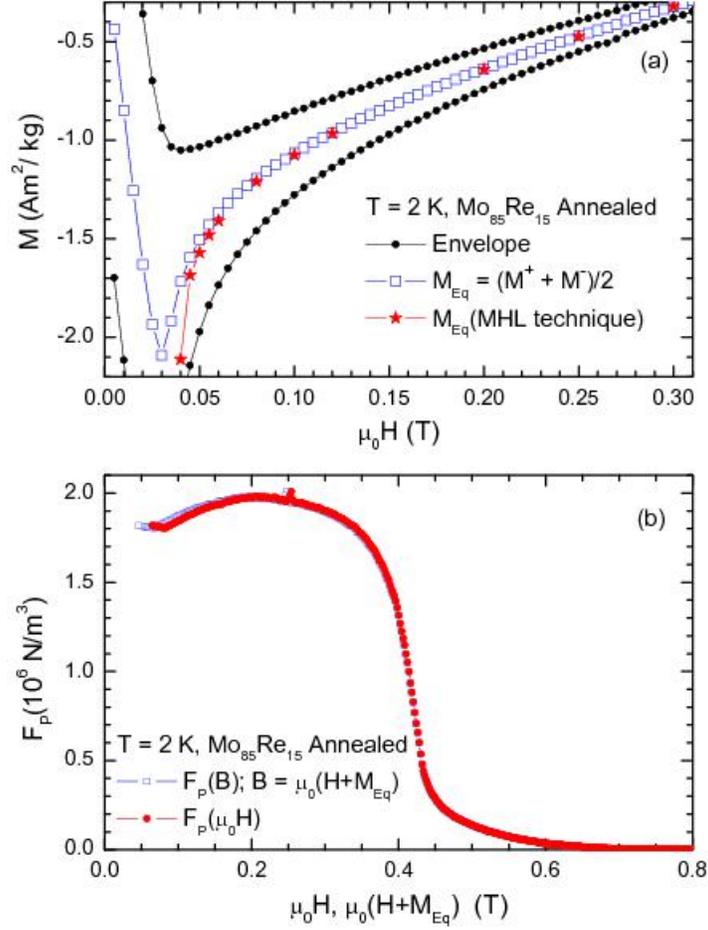

**FIG. 7(a).** Experimental determination (using two different methods) of equilibrium magnetization of the annealed $Mo_{85}Re_{15}$ alloy at 2 K. **(b).** The $F_p$ versus $\mu_0H$ and $F_p$ versus $B$ curves for the annealed $Mo_{85}Re_{15}$ alloy at 2 K.

A detailed analysis of $F_p(H)$ in terms of the size, spacing and nature of the pinning centres, and the nature of their interaction with the flux lines has been done by Dew-Hughes [27], and we have presented a similar analysis in a recent work [21]. To investigate into the flux line pinning mechanisms in the present alloys using Dew-Hughes model, we have used a normalized field $h$, where $h = \frac{\mu_0 H}{\mu_0 H^*}$, $H^*$ being the field at which the (approximately) linearly falling portion of the $F_p$ versus $\mu_0H$ curves (in the higher field side) extrapolate to $F_p = 0$ [24, 28]. We have found that the $F_p(h)$ curves for the present alloys cannot be fitted by any of the functions of form $F_p \propto h^p(1-h)^q$, where the values of $p$ and $q$ depend on the details of the pinning mechanism, as prescribed by Dew-Hughes [24, 27]. Qualitatively, from the $h$ value corresponding to the peak in $F_p$, it appears that the grain boundaries might be the major flux line pinning centres in the annealed and as-cast $Mo_{52}Re_{48}$ alloys (multi-phase). However, apart from the Dew-Hughes model the field dependence of pinning force density can also be analyzed with the help of Kramer's theory [29], where it is assumed that $F_p(h)$ is represented by the functions $\mathcal{F}_p(h)$ and $\mathcal{F}_s(h)$ respectively in low and high reduced fields. Here $\mathcal{F}_p(h)$ is a pinning force computed assuming that at low reduced fields the maximum pinning force for some of the crystal defects can be exceeded by the Lorentz force, so that the flux lines may be de-pinned from these pinning centres. On the other hand, $\mathcal{F}_s(h)$ is computed assuming that there are some strong pinning centres for whom the pinning force cannot be exceeded by the Lorentz force and the flux lines will rather shear plastically around these pinning centres. Thus, $\mathcal{F}_p(h)$ and $\mathcal{F}_s(h)$ are increasing and decreasing functions of $h$ respectively, and the peak in $F_p(h)$ is reached when $\mathcal{F}_p(h) = \mathcal{F}_s(h)$ [29]. In this model, the effective interaction of the flux lines with the pinning centres is not only a function of the distribution of the pinning centres in the material, but also of the shear strength of the flux lines (if the flux lines are not flexible, the pinning interactions will mostly cancel each other) [29- 31]. The probability $p(K_P)$ that a particular flux line will be pinned with a pinning strength $K_P$ is given by a distribution (around an average pinning strength $\langle K_P \rangle$) function. Various distribution functions, e.g., the Gaussian, and the single and double Poisson distribution functions have been assumed for $p(K_P)$ for the computation of $F_p(h)$ [29, 32]. We have found that none of these distribution functions provides complete explanation for the experimentally obtained $F_p(h)$ curves for the present $Mo_{100-x}Re_x$ alloys for the entire $h$ regime. However, the experimental $F_p(h)$ curves for the present $Mo_{100-x}Re_x$ alloys can still be explained over a large $h$ regime by using a double Poisson distribution function of the form [29],

$$p(K_P) = C_1 exp\left(-\frac{K_P}{\langle K_P \rangle_1}\right) + C_2 exp\left(-\frac{K_P}{\langle K_P \rangle_2}\right),$$

where, $C_1 \langle K_P \rangle_1 + C_2 \langle K_P \rangle_2 = 1$, and $\langle K_P \rangle = C_1 \langle K_P \rangle_1^2 + C_2 \langle K_P \rangle_2^2$ (1)

The double Poisson distribution function indicates that at least two types of flux line pinning centres having different pinning strength distributions are present in the present alloys. Using equation (1), the pinning force density is expressed as [29],

$$\begin{aligned}F_p(h) &= \mathcal{F}_p(h) + \mathcal{F}_s(h) \\ &= \frac{h^{1/2}}{(1-h)^2}\left\{C_1\langle K_P\rangle_1^2\left[1-\left(\frac{K_{Pm}}{\langle K_P\rangle_1}+1\right)exp\left(-\frac{K_{Pm}}{\langle K_P\rangle_1}\right)\right]\right. \\ &\quad \left.+ C_2\langle K_P\rangle_2^2\left[1-\left(\frac{K_{Pm}}{\langle K_P\rangle_2}+1\right)exp\left(-\frac{K_{Pm}}{\langle K_P\rangle_2}\right)\right]\right\} \\ &\quad + K_s h^{1/2}(1-h)^2 exp\left(-\frac{K_{Pm}}{\langle K_P\rangle}\right) \end{aligned} \quad (2)$$

Here $K_s$ is the shear strength parameter, and $K_{Pm} = K_s(1-h)^4$. The parameter $K_s$ mostly depends on the thermodynamic and upper critical fields of a superconductor, and depends only very weakly on the defect structure of a material. Therefore while fitting equation (2) to our experimental data we have assumed on the basis of Kramer's paper [29] that $\langle K_P\rangle_1 = m_1 K_s^{1/2}$, $\langle K_P\rangle_2 = m_2 K_s^{1/2}$, and $\langle K_P\rangle = mK_s$, where $m_1$, $m_2$ and $m$ are positive fractional numbers used as the fitting parameters along with $C_1$, $C_2$, and $K_s$. For equations (1) and (2), the reduced field $h$ was defined as $h = \frac{\mu_0 H}{\mu_0 H_{irr}}$. The irreversibility field $H_{irr}$ was determined from the $M(H)$ curves, where it is identified as the field at which the magnetic irreversibility $\Delta M$ goes to zero (within the resolution of the present experimental setup). In the case of the $Mo_{100-x}Re_x$ alloys, there is some scope of uncertainty in the determination of $H_{irr}$ because of the presence of a long tailing hysteresis in $M(H)$ (as discussed earlier). To avoid this uncertainty we have estimated $H_{irr}$ by extrapolating the $J_c(H)$ curves to $J_c = 0$ in the high $H$ side.

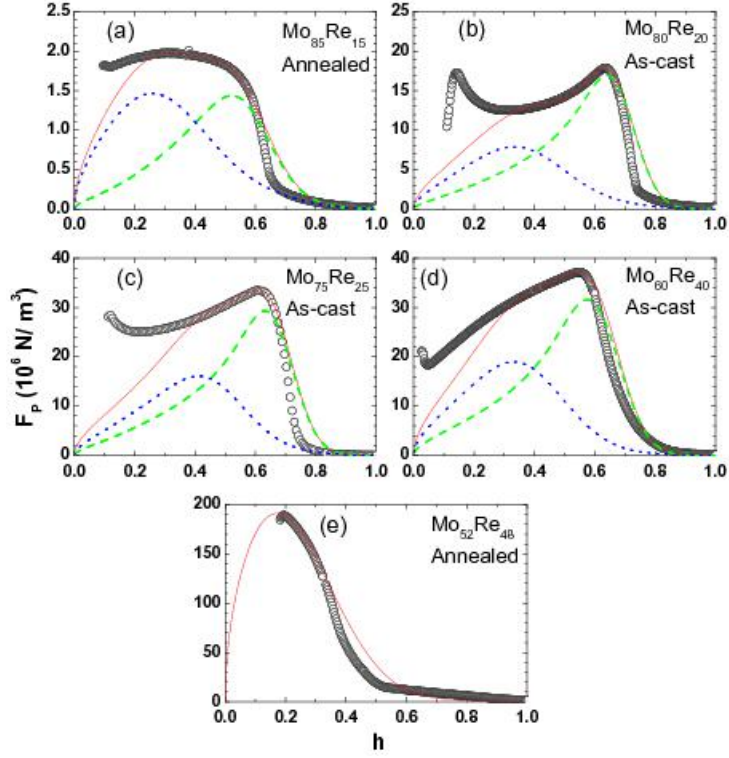

**FIG. 8.** The $F_p$ versus $h$ curves for the Mo$_{100-x}$Re$_x$ alloys at 2 K, where $h = \frac{H}{H_{irr}}$. The (black) open circles represent the experimental data and the (red) solid line represents the fitted curve [equation (2)]. The (blue) dotted line and the (green) dashed line respectively represent two components of the fitted (red) curve.

Fig. 8 shows the $F_p$ versus $h$ curves (where $h = \frac{H}{H_{irr}}$) for the present Mo$_{100-x}$Re$_x$ alloys at 2 K. In this figure, the (red) solid line represents the best fit curve obtained using equation (2). The (blue) dotted line and the (green) dashed line respectively represent the two components of this best fit (red) curve obtained by putting $C_1 = 0$ and $C_2 = 0$ alternately (keeping the other parameters unchanged). By correlating our $J_c(H)$, $F_p(H)$, and $F_p(h)$ curves (Figs. 5, 6 and 8) we find that the best fit curves in Fig. 8 explain the experimental $F_p(h)$ curves only in the "small bundle" flux line pinning regime. In higher $H$, the best fit curves lie above the experimentally obtained $F_p(h)$ curves, indicating a reduction of the effective pinning strength. We have found that in this high $H$ regime, the Mo$_{100-x}$Re$_x$ alloys exhibit strong signature of magnetic relaxation. Fig. 9 shows the time (*t*) dependence of $M/|M_0|$ (where $M_0$ is the magnetization at $t = 0$) in the $Mo_{75}Re_{25}$ alloy in three *H*-regimes on the $M^+$ curve. The highest *H*-value in fig. 9 corresponds to the regime where the best fit curve corresponding to equation (2) lie above the experimentally obtained $F_p(h)$ curves. The magnetic relaxation effect is not appreciable in the lower *H*-regimes. The *M(t)* curve obtained in the high *H*-regime can be fitted with the following the equation,

$$\frac{M(t)}{|M_0|} = 1 + \frac{k_B T}{U_0} ln\left(1 + \frac{t}{t_0}\right) \qquad (3)$$

where $U_0$ represents the activation energy. The fitting of this equation to the experimental curve indicates the presence of flux creep effects in the alloy in the high *H* regime. We have found that the corresponding *J_c(H)* curve for the $Mo_{75}Re_{25}$ alloy exhibits a rapid drop in this high *H* regime (see Fig. 5). Thus the reduction of the effective pinning strength and *J_c(H)* in high *H* is due to the flux creep effects. Qualitatively similar results were obtained for the other $Mo_{100-x}Re_x$ alloys as well.

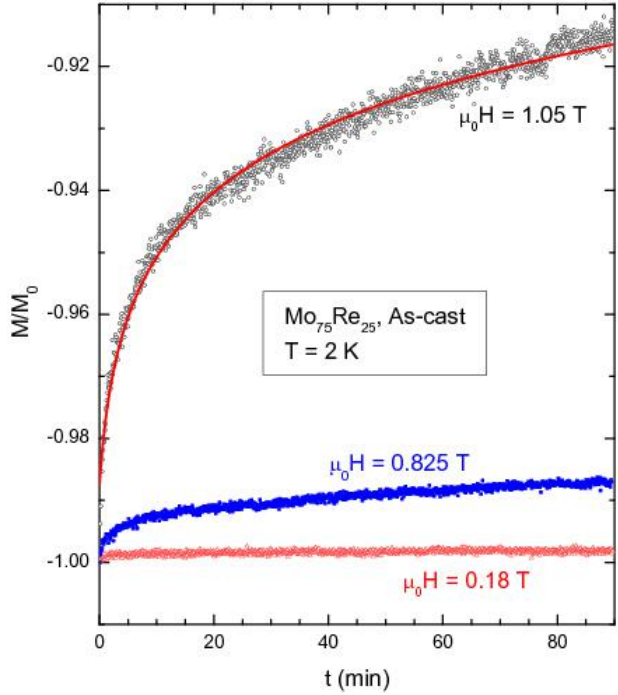

**FIG. 9.** Relaxation of magnetization in the $Mo_{75}Re_{25}$ alloy in different field regimes. These experiments were performed in MPMS-3 SQUID VSM.

From fig. 8 we find that for the annealed $Mo_{85}Re_{15}$ alloy and the as-cast $Mo_{80}Re_{20}$, $Mo_{75}Re_{25}$, and $Mo_{60}Re_{40}$ alloys [panels (a)- (d)], the two components of the fitted $F_p(h)$ curve exhibit peaks near $h = 0.3$ and 0.5, $h = 0.3$ and 0.6, $h = 0.4$ and 0.6, and $h = 0.3$ and 0.6 respectively. According to Kramer, a peak in $F_p(h)$ at a low value of $h$, e. g. close to 0.2, may be ascribed to flux line pinning by the grain boundaries present in a material [24, 33], and a peak in $F_p(h)$ at a high value of $h$ (between 0.7- 0.8) may be expected to be due to flux line pinning by dislocation networks [34]. Voids, on the other hand give rise to a peak in $F_p(h)$ at a value of $h$, between 0.3 and 0.7 [34]. Our optical metallography experiments show strong signatures of the presence of dislocation networks in the above mentioned alloys. In the sample characterization methods employed in the present work, however, we cannot see the evidence of the presence of voids or any other point pinning defect structure in our $Mo_{100-x}Re_x$ alloys, though it is reported that it is possible to create voids in the $Mo_{100-x}Re_x$ alloys by cold-work (e.g., rolling) [7]. Interstitial impurities like oxygen can also act as point pinning centres like the voids [34]. Though our samples have been synthesized in 99.99+ % Ar atmosphere (both during melting and annealing), a small amount of interstitial impurity cannot be completely ruled out. Voids and the interstitial defects mentioned above may be expected to have dimensions ~1 nm. The dislocations are visible in the present optical metallography because of the lining up of the etch pits. Actually, the width of the dislocations are expected to extend over a few atomic spacing [35], or in other words, over a few nm. Using the formula $\xi = \left(\frac{\varphi_0}{2\pi H_{c2}}\right)^{1/2}$, we have found that the coherence length of all the present alloys is close to 10 nm. Therefore, it is likely [12] that the flux lines in the present alloys may be pinned across the dislocations and the point defects mentioned above. It is also reported in literature that for the dislocation networks, the interaction with the flux lines is between the strain field of the dislocations and the stress field of the flux line lattice [36]. The interaction between the dislocation network and the flux lines in the present alloys will, therefore, depend on the size of the flux line bundles and the area of dislocation loops [36, 37]. A full proof first principle theory for such an interaction does not exist, and large pinning forces due to dislocation networks have been reported for $h = 0.4$ and 0.55 as well [36]. Accordingly, for the alloys mentioned above we believe that the peak in the component of $F_p(h)$ at $h$ = 0.3- 0.4 is due to the point defects mentioned above, and the peak at $h$ = 0.5- 0.6 is due to the pinning of flux lines by the dislocation networks. The grain boundaries do not seem to have a major role in pinning the flux lines in the single phase $Mo_{100-x}Re_x$ alloys in the above mentioned field regimes. Fig. 8(e) shows the experimental and best-fit $F_p(h)$ curves for the annealed $Mo_{52}Re_{48}$ alloy at 2 K. Though much of the data had to be excluded because of the flux jump effects, it may still be observed in fig. 8(e) that the best-fit curve representing equation (2) has only one component in this case ($C_1$ = 0), and it peaks close to $h$ = 0.2. Qualitatively similar fitted curve is obtained for the as-cast $Mo_{52}Re_{48}$ alloy as well, indicating flux line pinning by the grain boundaries in these alloys [29, 33]. It may be recalled that we had previously reached the same conclusions about the flux line pinning mechanism in the as-cast and annealed $Mo_{52}Re_{48}$ alloys, using the Dew-Hughes model [27]. Thus, in spite of the presence of the $\sigma$ and $\chi$ phases, we do not find any additional flux line pinning mechanism in these multi-phase alloys. It may be noted in this context that the $\sigma$ and $\chi$ phases in the $Mo_{100-x}Re_x$ alloys are superconducting at low temperatures [11]. Any additional superconducting phase in a sample acts as a flux line pinning centre when its $\xi$ value is different from the major phase [27]. The superconducting properties of the $\sigma$ phase $Mo_{100-x}Re_x$ alloys are already reported in literature. From the reported $H_{c2}$ values [11], we have estimated the $\xi$ of these alloys using the method described above. We have found that the $\xi$ values for the $\sigma$ and $\beta$ phase $Mo_{100-x}Re_x$ alloys are approximately equal (close to 10 nm). We could not find literature reports on the $H_{c2}$ or $\xi$ of the $\chi$ phase $Mo_{100-x}Re_x$ alloys. However, in view of the present results we believe that even for the $\chi$ phase, the $\xi$ value is not significantly different from that of the $\beta$ phase alloys, and thus it is possible that there is only one major flux line pinning mechanism in the as-cast and annealed $Mo_{52}Re_{48}$ alloys.

We have already explained that the $F_p(H)$ curves (fig. 6) for the single phase $Mo_{100-x}Re_x$ alloys indicate the presence of a peak in very low $H$. Comparing figs. 6 and 8 it is clear that this peak is present close to $h$ = 0.1 or even below. A peak in the $F_P(h)$ curves at such a low value of $h$ is not normally expected in the existing theories [27, 29, 33] of flux line pinning. Literature reports suggest that such a low *h* peak may arise due to a large anisotropy in the upper critical field, e.g., in the case when the ratio of the upper critical fields along the *ab*-plane and the *c*-axis is nearly 5 [38, 39]. Such high anisotropy, however, cannot be expected for the present system of alloys- especially in the single (*β*) phase compositions where the indication of this peak is observed. On the other hand, we have recently found strong signature of the existence of two unequal superconducting gaps in the single phase $Mo_{100-x}Re_x$ alloys [9]. Our results suggest that the Re-5*d* like states at the Fermi level do not intermix with the Mo-5*p* and 5*s* states in the $Mo_{100-x}Re_x$ alloys, and this leads to the two-gap nature of these superconductors [9]. Previously, we have also shown in the case of $PrPt_4Ge_{12}$ that a peak in the $F_p(h)$ curves near $h = 0.1$ may be related to the existence of two superconducting gaps in that material [38]. Similar results also exist in the case of $MgB_2$, which is another system with two superconducting energy gaps [40- 42]. The isothermal $J_c(H)$ of $MgB_2$ was found to follow a two-exponential model, and this model was successfully used to numerically reproduce the experimental *M(H)* curves. The two exponential terms in the $J_c(H)$ were thought to arise due to the two superconducting energy gaps in $MgB_2$ [41, 42]. Using the same model the peak in the $F_p(h)$ curves near $h = 0.1$ was also explained with appreciable quantitative accuracy in the case of $MgB_2$. It appeared that the peak in $F_p(h)$ arising due to grain boundary pinning shifts to lower *h* values because of the presence of the two superconducting energy gaps [41, 42]. We found that the same model works for $PrPt_4Ge_{12}$ as well [38]. We therefore tried to fit the same model in the case of the present alloys. Fig. 10(a) shows the $J_c(H)$ curve of the as-cast $Mo_{60}Re_{40}$ alloy at 2 K, where the low *H* data have been fitted by an expression of the form $J_c = A exp(-H/H_{c2}) + B exp(-H/H_{c2})$. Since the fitting was found to be quite good in low *H*, the fitted curve was used to generate the corresponding $F_p(h)$ curve for this alloy. Fig. 10(b) shows the $F_p(h)$ curve for the as-cast $Mo_{60}Re_{40}$ alloy, where the thin (red) solid line represents equation (2), and the dashed (blue) line represents the $F_p(h)$ generated using the two-exponential model. It is observed that the experimental $F_p(h)$ curve in low *H* is almost reproduced by the model mentioned above in the *H* regime where equation (2) cannot fit the experimental data. The two-exponential model, however, does not fit the experimental $J_c(H)$ curves in the higher $H$ regime [see figs. 5 and 10(a)] and the $F_p(h)$ generated using the two-exponential model does not fit the experimental $F_p(h)$ curve in this regime. Qualitatively similar results exist for the other single ($\beta$) phase Mo$_{100-x}$Re$_x$ alloys as well. We have earlier pointed out that the $J_c(H)$ curves in the Mo$_{100-x}$Re$_x$ alloys exhibit a clear departure from the expected [23] behaviour in the low $H$ regime where the $J_c$ decreases very sharply with increasing $H$ over quite a broad $H$ regime. The analysis presented above indicates that this behaviour is because of the presence of two superconducting energy gaps in these alloys [9]. The smaller of these two energy gaps in the Mo$_{100-x}$Re$_x$ alloys probably does not exist in higher $H$. In this regime the $F_p(h)$ behaviour is reasonably explained by equation (2), while this equation fails in the presence of two gaps in the low $H$ regime. We have mentioned earlier that none of the present alloys provide any indication of the presence of weak links, and this shows that the width of the grain boundaries in these alloys is less than the $\xi$ [43]. In such a case the self-energy of a flux line is reduced at the grain boundaries, and this gives rise to a pinning force [44-46]. The grain boundaries are also known to contain dislocations [46, 47] and thus become effective flux line pinning centres. We also recall that the Mo$_{100-x}$Re$_x$ alloys with the smaller grain size exhibit the higher $J_c$ (which is in the low $H$ regime) in the present study, indicating that grain boundaries are capable of pinning the flux lines in the Mo$_{100-x}$Re$_x$ alloys. Hence we find it reasonable to believe that grain boundary pinning is present even in the $\beta$ phase Mo$_{100-x}$Re$_x$ alloys, though the fitting of equation (2) does not reveal the same. We therefore argue that the $F_p(h)$ behaviour of the single phase Mo$_{100-x}$Re$_x$ alloys in the low $H$ regime is probably due to grain boundary pinning modified by the presence of two superconducting gaps [41, 42]. In this logic the asymmetry of the isothermal $M(H)$ curves in Fig. 3 may be ascribed to grain boundary pinning modified by the presence of two superconducting energy gaps in the Mo$_{100-x}$Re$_x$ alloys, along with the overall low pinning strength in different $H$ regimes. The features related to this two gap phenomenon are not really observed in the multi-phase Mo$_{100-x}$Re$_x$ alloys. The presence of additional ($\sigma$ and $\chi$) phases with different crystal structures probably suppress the signatures of the presence of two superconducting gaps.

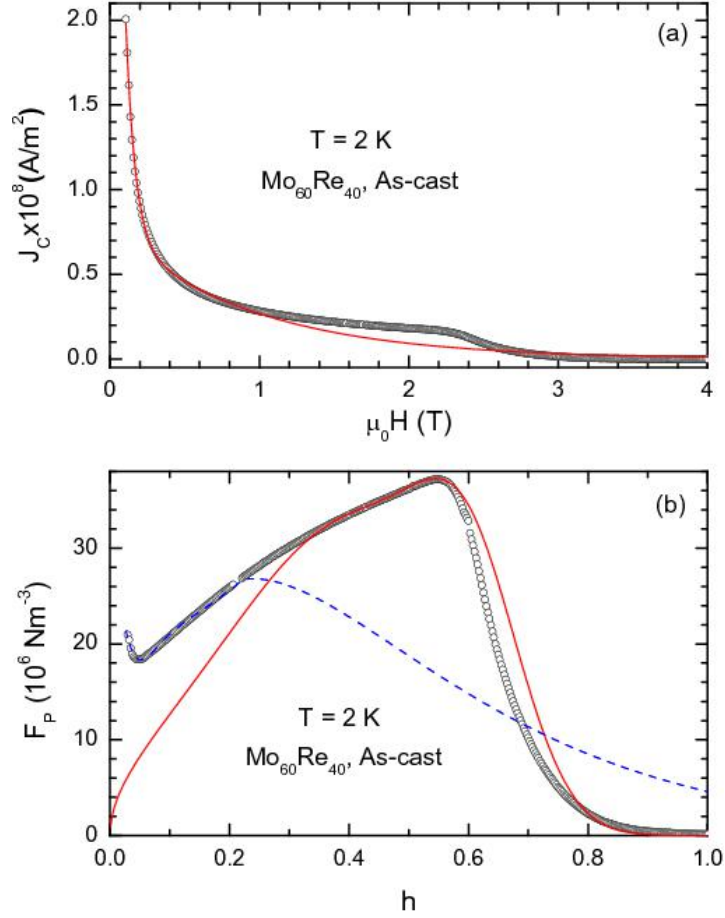

**FIG. 10(a).** Field dependence of $J_c$ of the as-cast $Mo_{60}Re_{40}$ alloy, where the low field data has been fitted by a two-exponential model. **(b)** The $F_p$ versus $h$ curve for the as-cast $Mo_{60}Re_{40}$ alloy. The (black) open circles represent the experimental data, the (red) solid line represents the best-fit curve obtained using equation (2), and the dashed (blue) line represents the $F_P(h)$ values generated using the two-exponential model mentioned above.

## 4. Summary and conclusions

Polycrystalline samples of $Mo_{100-x}Re_x$ ($15 \leq x \leq 48$) alloys were synthesized using the arc melting technique, and the samples were characterized using XRD, optical metallography, and EDX analysis. Single $\beta$ phase alloys and with $x \leq 40$ compositions, and alloys with $x \geq 40$ containing multiple phases ($\beta$, σ and χ phases) were identified. The grain size in the $Mo_{100-x}Re_x$ alloys was found to be varying from ~100 μm to mm level, and the multi-phase alloys were found to have smaller grains. Optical micrographs showed the abundance of dislocation networks in the *β* phase alloys. The isothermal field dependence of magnetization of most of the alloys revealed a pronounced asymmetry between the field increasing and field decreasing curves. Experimental studies revealed that this asymmetry is not related to the surface barrier effects. The critical current density and the pinning force density of the alloys were estimated from the field dependence of magnetization. Analysis using existing theories indicated that the dislocation networks and point pinning centers like voids and interstitial imperfections are the main flux line pinners in the $Mo_{100-x}Re_x$ alloys in the "small-bundle" flux line pinning regime. The critical current density is also found to be relatively robust against increasing magnetic field in this "small-bundle" regime. In still higher fields, the critical current density is suppressed by flux creep. In the low field regime, on the other hand, the dislocation networks and the point pinning centres mentioned above are not very effective in pinning the flux lines. Contrary to a previous report [8], the grain boundaries appear to pin the flux lines in this regime, and in fact, alloys with smaller grain size are found to exhibit higher critical current density in the low field regime. This field dependences of the critical current and the pinning force density of the present alloys in the low field regime were analyzed in the light of our finding that the $Mo_{100-x}Re_x$ alloys are a two gap superconducting system [9]. Our results indicate that the presence of two superconducting energy gaps in the present system modifies the field dependence of critical current density, and this produces a low field maximum in the field dependence of pinning force density. The asymmetry in the isothermal field dependence of magnetization seems to be related to the effect of the presence of two superconducting energy gaps on the grain boundary pinning mechanism in low fields along with the overall low flux line pinning strength in the higher field regimes.

## Acknowledgement

We thank Shri R. K. Meena for his help in sample preparation, and Dr. Gurvinderjit Singh for the XRD measurements. We also thank Shri Rakesh Kaul and Dr. M. A. Manekar for their help in optical metallography.

## References

[1] S. V. Vonsovsky, Yu. A. Izyumov and E. Z. Kurmaev, *Superconductivity of Transition Metals: Their Alloys and compounds*, 1982, Translated by Brandt E H and Zavarnitsyn A P, Springer Verlag, Berlin (and references therein); T. A. Ignat'eva and Yu. A. Cherevan', Pis'ma Zh. Eksp. Teor. Fiz. **31** (1980) 389.

[2] J. Wardsworth and J. P. Wittenauer, *Evolution of Refractory Metals and Alloys*, 1993, editors: E. N. C. Dalder, T. Grobstein and C. S. Olson, The Minerals, Metals and Materials Society, Warrendale, OH, p. 85; R. L. Heenstand, *Evolution of Refractory Metals and Alloys*, 1993, editors: E. N. C. Dalder, T. Grobstein and C. S. Olson, (The Minerals, Metals and Materials Society, Warrendale, OH., p 109.

[3] R. L. Mannheim and J.L Garin, J. Mater. Process. Technol. **143-144** (2003) 533; S. R. Agnew and T. Leonhardt, JOM: The Journal of the Minerals, Metals & Materials Society **55** (2003) 25.

[4] P. Mao, K. Han and Y. Xin, J. Alloys and Comp. **464** (2008) 190.

[5] J. E. Kunzler, Rev. Mod. Phys. **33** (1961) 501; J. E. Kunzler, E. Buehler, F. S. L. Hsu, B. T. Matthias and C. Wahl, J. Appl. Phys. **32** (1961) 325.

[6] M. J. Witcomb, A. Echarri, A. V. Narlikar and D. Dew-Hughes, J. Materials Sci. **3** (1968) 191.

[7] M. J. Witcomb and D. Dew-Hughes, Acta Metallurgica **20** (1972) 819.

[8] M. J. Witcomb and D. Dew-Hughes, J. Less-Comm. Met. **31** (1973) 197.

[9] S. Sundar, L. S. Sharath Chandra, M. K. Chattopadhyay and S. B. Roy, J. Phys.: Condens. Matter **27** (2015) 045701.

[10] E. Lerner and J. G. Daunt, Phys. Rev. **142** (1966) 251.

[11] F. J. Morin and J. P. Maita, Phys. Rev. **129** (1963) 1115; H. R. Khan and C. J. Raub, J. Less-Comm. Met. **69** (1980) 361; C. C. Koch and J. Q. Scarbrough, Phys. Rev B **3** (1971) 742.

[12] H. Ullmaier, *Irreversible Properties of Type II Superconductors*, 1975, Springer-Verlag, Berlin Heidelberg.

[13] C. P. Bean and J. D. Livingstone, Phys. Rev. Lett. **12** (1964) 14.

[14] L. Burlachkov, M. Konczykowski, Y. Yeshurun and F. Holtzberg J. Appl. Phys. **70** (1991) 5759; L. Burlachkov, Phys. Rev. B **47** (1993) 8056.

[15] Y. Matsumoto, T. Akune and N. Sakamoto, J. Phys.: Conference Series **150** (2009) 052157.

[16] M. Zehetmayer, M. Eisterer, J. Jun, S. M. Kazakov, J. Karpinski, A. Wisniewski and H. W. Weber, Phys. Rev. B **66** (2002) 052505.

[17] C. P. Bean, Rev. Mod. Phys. **36** (1964) 31.

[18] D. N. Zheng, H. D. Ramsbottom and D. P. Hampshire, Phys. Rev. B **52** (1995) 12931.

[19] E. Martínez, P. Mikheenko, M. Martínez-López, A. Millán, A. Bevan and J. S. Abell, Phys. Rev. B **75** (2007) 134515.

[20] M. Tinkham, Helv. Physica Acta **61** (1988) 443.

[21] Md. Matin, L. S. Sharath Chandra, M. K. Chattopadhyay, R. K. Meena, Rakesh Kaul, M. N. Singh, A. K. Sinha and S. B. Roy, J. Appl. Phys. **113** (2013) 163903.

[22] S. Chaudhary, S. B. Roy, P. Chaddah, P. K. Babu, R. Nagarajan and L. C. Gupta, Phil. Mag. B **80** (2000) 1393.

[23] G. Blatter, M. V. Feigel'man, V. B. Geshkenbein, A. I. Larkin and V. M. Vinokur, Rev. Mod. Phys. **66** (1994) 1125.

[24] J. W. Ekin, Supercond. Sci. Technol. **23** (2012) 083001; and references therein.

[25] M. Zehetmayer, Phys. Rev. B **80** (2009) 104512.

[26] P. Chaddah, S. B. Roy and M. Chandran, Phys. Rev. **59** (1999) 8440.

[27] D. Dew-Hughes, Phil. Mag. **30** (1974) 293.

[28] V. Sandu, Mod. Phys. Lett. B **26** (2012) 1230007.

[29] E. J. Kramer, J. Appl. Phys. **44** (1973) 1360.

[30] W. A. Fietz and W. W. Webb, Phys. Rev. **178** (1969) 657.

[31] R. Labusch, Cryst. Lattice Defects **1** (1969) 1.

[32] S. Kumar, S. N. Kaul, J. R. Fernandez and L. F. Barquin, Phys. Lett. A **374** (2009) 335.

[33] E. J. Kramer and G. S. Knapp, J. Appl. Phys. **46** (1975) 4595.

[34] E. J. Kramer, Nuclear Mater. **72** (1978) 5.

[35] W. D. Callister, Jr., *Materials science and engineering: an introduction*, 2006, John Wiley, Singapore.

[36] E. J. Kramer, J. Appl. Phys. **49** (1978) 742.

[37] E. J. Kramer, Phil. Mag. **33** (1976) 331.

[38] L. S. Sharath Chandra, M. K. Chattopadhyay and S. B. Roy, Supercond. Sci. Technol. **25** (2012) 105009.

[39] M. Eisterer, Phys. Rev. B **77** (2008) 144524.

[40] S. Tsuda, T. Yokoya, Y. Takano, H. Kito, A. Matsushita, F. Yin, J. Itoh, H. Harima and S. Shin, Phys. Rev. Lett. **91** (2003) 127001.

[41] Z. X. Shi, J. Wang, H. Lv and T. Tamegai, Physica C **449** (2006) 104.

[42] J. Wang, Z. X. Shi, H. Lv and T. Tamegai, Physica C **445-448** (2006) 462.

[43] A. Gupta, M. Decroux, P. Selvam, D. Cattani, T. C. Willis and Ø. Fischer, Physica C **234** (1994) 219.

[44] G. Zerweck, J. Low Temp. Phys. **42** (1981) 1.

[45] D. Dew-Hughes, Phil. Mag. B **55** (1987) 459.

[46] A. Das Gupta, C. C. Koch, D. M. Kroeger and Y. T. Chou, *How effectively can grain boundaries pin flux lines*, Advances in Cryogenic Engineering, vol. 24, 1978, editor: K. D. Timmerhaus, Plenum Press, New York, p. 350.

[47] P. Mikheenko, E. Martinez, A. Bevan, J. S. Abell and J. L. MacManus-Driscoll, Supercond. Sci. Technol. **20** (2007) S264.